\documentclass[a4paper]{article}

\usepackage{INTERSPEECH2021}
\usepackage{multirow}
\usepackage{lineno}
\usepackage{makecell}

\title{Investigation of Practical Aspects of Single Channel \\ Speech Separation for ASR}
\name{Jian Wu$^1$, Zhuo Chen$^1$, Sanyuan Chen$^2$, Yu Wu$^1$, Takuya Yoshioka$^1$, \\ Naoyuki Kanda$^1$,  Shujie Liu$^1$, Jinyu Li$^1$}
\address{
$^1$Microsoft Corporation ~~~~~
$^2$Harbin Institute of Technology}
	
\email{ 
\begin{tabular}{c} 
\{wujian, zhuc, yuwu1, tayoshio, nakanda, shujliu, jinyli\}@microsoft.com, \\
sychen@ir.hit.edu.cn
\end{tabular}
}

\begin{document}

\maketitle
\begin{abstract}

Speech separation has been successfully applied as a front-end processing module of conversation transcription systems thanks to its ability to handle overlapped speech and its flexibility to combine with downstream tasks such as automatic speech recognition (ASR). 
However, a speech separation model often introduces target speech distortion, resulting in a sub-optimum word error rate (WER). 
In this paper, we describe our efforts to improve the performance of a single channel speech separation system. Specifically, we investigate a two-stage training scheme that firstly applies a feature level optimization criterion for pre-training, followed by an ASR-oriented optimization criterion using an end-to-end (E2E) speech recognition model.
Meanwhile, to keep the model light-weight, we introduce a modified teacher-student learning technique for model compression. By combining those approaches, we achieve a absolute average WER improvement of $2.70\%$ and $0.77\%$ using models with less than 10M parameters compared with the previous state-of-the-art results on the LibriCSS dataset for utterance-wise evaluation and continuous evaluation, respectively.
\end{abstract}

\section{Introduction}

Deep learning based speech separation has been investigated in recent years since the proposal of deep clustering (DPCL) \cite{hershey2016deep} and permutation invariant training (PIT) \cite{kolbaek2017multitalker}. 
Various follow-up studies have been reported, including exploration of the different architectures \cite{luo2018speaker,wang2018alternative,luo2019conv} and recipes \cite{li2019listening,takahashi2019recursive,ge2020spex+}, extension to multi-channel processing \cite{luo2020end,wang2018combining} and joint modeling with other tasks such as automatic speech recognition (ASR) \cite{settle2018end,chang2019end}. These advances resulted in improved performance with respect to several metrics related to audio quality and ASR accuracy.
% Recently, the CHiME6 challenge \cite{watanabe2020chime} and LibriCSS dataset \cite{chen2020continuous} bring the separation technology to the real world task and give higher priority to the ASR evaluation results.
% yoshioka2019advance raj2021integration

% the improvements on separation quality, 
Recently, the multi-channel speech separation achieves good performance \cite{kanda2019guided,wu2020end} and  has been successfully integrated into conversation transcription systems \cite{yoshioka2019advances}. However, the improvement has still been limited with single channel input for the conversational tasks \cite{chen2020continuous,wang2020multi,chen2020conformer}.
The single channel conversation transcription remains challenging for two reasons. Firstly, a single channel network can not be benefited from the spatial information, leading to inferior separation.
Secondly, the single channel separation often
employs a signal-level objective function, which is known to introduce speech distortion that hurts the accuracy of ASR systems \cite{chen2018building}.

% \cite{chen2020continuous,wang2020multi,chen2020conformer} due to the inaccessibility of spatial feature\cite{yoshioka2018multi} and beamforming.

To overcome this limitation, several methods have been studied. 
In \cite{wang2019enhanced,wang2020voicefilter}, a feature-level objective function was found to be beneficial for ASR. The joint optimization of the front-end with the back-end ASR was also shown to be effective in both speech enhancement \cite{gao2015joint,wang2016joint,menne2019investigation} and separation \cite{von2020end,von2020multi}. However, despite these promising results, the past studies on ASR-oriented speech separation lacks consideration of two aspects.  Firstly, the ASR back-end used in the past works were either jointly updated or tightly connected to the separation model, which could cause undesirable word error rate (WER) degradation on out-of-domain dataset and less flexibility of using each module in different application scenarios. Secondly, a jointly trained speech separation model may suffer from performance degradation when the back-end ASR model is changed.

In this paper, we attempt to answer the above questions experimentally by using the LibriCSS dataset with the goal of improving the performance of the single channel speech separation for conversation transcription. We adopt a training recipe which can leverage knowledge from both an ASR system and spectrum reconstruction. Specifically, our training recipe contains two stages, where a seed model is firstly trained under a conventional mask-based feature approximation objective, followed by fine-tuning with the ASR-oriented training criteria using an end-to-end ASR network. The ASR model used for training could be different from that employed for evaluation, and its parameters are kept fixed during training to prevent the co-adaptation of the front-end and ASR model. In addition, our comparison shows that a Conformer-based separation model outperforms several other state-of-the-art model architectures. As regards the model architecture, we also examine the effect of model compression using the layer-wise teacher-student learning proposed recently \cite{sun2019patient} to achieve fast inference and lower runtime cost. Finally, we report results for both the utterance-wise and continuous evaluation settings of LibriCSS, outperforming the previously reported best numbers. 

\section{System Description}

% In this section, we will briefly overview the signal model as well as the primary network architecture adopted for single channel separation. Then the procedure of the two-stage training scheme and teacher-student learning is introduced.

\subsection{Signal Model}
We consider the following signal model with $C$ speakers. The single channel far-field signal $\mathbf{y}$ in time domain is impaired by reverberation and additive noise:
\begin{equation}
    \mathbf{y} = \sum_c \mathbf{x}_c + \mathbf{n} = \sum_c \mathbf{s}_c * \mathbf{h}_c + \mathbf{n}, \label{eq:signal_model}
\end{equation}
where $\mathbf{h}_c$ is room impulse response (RIR) between speaker c ($0 \leqslant c < C$) and the microphone. $\mathbf{s}_c$ is c-th source speaker and $\mathbf{x}_c$ denotes the corresponding image signal. We model environment noise as $\mathbf{n}$, consisting of directional and isotropic noise. After applying short-time Fourier transform (STFT), $(1)$ is converted into frequency domain:
\begin{equation}
    \mathbf{Y} = \sum_c \mathbf{X}_c + \mathbf{N},
\end{equation}
where $\{\mathbf{Y}, \mathbf{N}, \mathbf{X}_c\} \in \mathbb{C}^{T \times F}$. $T$ and $F$ denote the total number of time frames and frequency bins.

\subsection{Conformer Structure}

We employ the frequency domain model $\mathcal{M}(\cdot)$ for single channel separation as it was found to be more robust for speech recognition than time domain structures from our preliminary results. The network is trained to estimate the time-frequency masks (TF-masks) on image signals, i.e., $\mathbf{M}_{0,1} = \mathcal{M}(\mathbf{Y}) \in \mathbb{R}^{T \times F}$. Inverse STFT (iSTFT) is used for signal reconstruction for the additional post processing
\begin{equation}
    \mathbf{x}'_c = \text{iSTFT}(\mathbf{Y} \odot \mathbf{M}_c).
\end{equation}

Following the work in \cite{chen2020conformer}, a modified Conformer \cite{gulati2020conformer} structure that consists a stack of conformer blocks is used as $\mathcal{M}(\cdot)$ for TF-masks estimation. Our conformer block is composed of three parts, i.e., a multi-head self-attention module (MHSA) \cite{vaswani2017attention}, a convolution module (CONV) and a feedforward network (FFN). The output of the conformer block $z_3$ is calculated as the following equations given the input $z_0$
\begin{equation}
\begin{aligned}
    z_1 & = z_0 + \text{MHSA}(\text{ln}(z_0)) \\
    z_2 & = z_1 + \text{CONV}(\text{ln}(z_1)) \\
    z_3 & = z_2 + \text{FFN}(\text{ln}(z_2))
\end{aligned}
\end{equation}
where $\text{ln}(\cdot)$ denotes the layer normalization. The dropout layers within the conformer block are disabled.

For the implementation of the MHSA, we adopt the learnt relative position encoding described in \cite{shaw2018self} instead of the original version \cite{dai2019transformer}. Besides, a squeeze-and-excitation (SE) \cite{hu2018squeeze} block is added to the last layer of the convolution module which operates on the output of the second pointwise convolution.

% The convolution module starts with a pointwise convolution and a gated linear unit (GLU), followed by a 1-D depthwise convolution layer, a batch normalization layer, a Swish activation and a second pointwise convolution.
\subsection{Network Training}

Although the prior work in \cite{chen2020conformer} has reported significant improvement in terms of WER by using the Conformer structure for speech separation, a large performance gap is observed between the single channel and multi-channel systems, i.e., $5.9\%$ and $8.2\%$  absolute WER increase when overlap ratio is $30\%$ and $40\%$. 
One of the potential reasons lies in the mismatch between the front-end objective function and speech recognition. In a typical mask learning based separation network,
ideal amplitude mask (IAM) is used as the training target and the loss function for spectrum approximation is defined as
\begin{equation}
    \mathcal{L}_\text{SA} = \arg \min_{\phi \in \mathcal{P}} \sum_{(i, j) \in \phi} \left \Vert \mathbf{M}_i \odot |\mathbf{Y}| - |\mathbf{X}_j| \right \Vert_F
\end{equation}
under the permutation invariant training criteria. $\mathcal{P}$ refers all the possible permutations over $C = 2$ speakers and $\Vert \cdot \Vert_F$ denotes the Frobenius norm. 

However, the modern ASR model uses the acoustic features such as filter-bank (fbank), MFCC as the network input, which employs a representation with considerably lower dimension than acoustic mask. 
Thus the spectrum approximation in front-end may be redundant for the reconstruction of the acoustic features, which not only increases the difficulties of the network optimization, but also puts mismatched emphasis for signal recovery.
For example, the fbank feature gives more weights on the lower frequency regions, while the IAM treats each frequency equally.

% followed \cite{wang2019enhanced}
To lead the better estimation of the acoustic features that the ASR back-end expected and reduce the potential feature distortion, we modified equation $(5)$ to measure the feature level difference instead of the amplitude:
\begin{equation}
    \mathcal{L}_\text{FA} = \arg \min_{\phi \in \mathcal{P}} \sum_{(i, j) \in \phi} \left \Vert \mathcal{F}(\mathbf{M}_i \odot |\mathbf{Y}|) - \mathcal{F}(|\mathbf{X}_j|) \right \Vert_F
\end{equation}
where $\mathcal{F}(\cdot)$ denotes the feature transform function. In order to further match the ASR input, on top of the model optimized with $(6)$, we adopt a well trained ASR model and tune the separation model using ASR's training criteria. In this work, during training,  we employ encoder-decoder based E2E speech recognition model in the experiments, with a hybrid CTC \& attention objective function \cite{watanabe2017hybrid}:

\begin{equation}
    \mathcal{L}_\text{ASR} = \sum_c \lambda \log p_\text{ctc}(\mathbf{r}_c|\mathbf{X}'_c) + (1 - \lambda) \log p_\text{dec}(\mathbf{r}_c|\mathbf{X}'_c) \label{eq:asr_loss}
\end{equation}
where $\mathbf{X}'_c = \mathcal{F}(\mathbf{M}_c \odot |\mathbf{Y}|)$ denotes the feature sequence of the separated speaker $c$ and $\mathbf{r}_c$ is the corresponding transcription. $\lambda$ is a hyper parameter to balance the CTC loss on encoder branch $\log p_{\text{ctc}}(\cdot)$ and the cross-entropy loss $\log p_{\text{dec}}(\cdot)$ on the decoder predictions. 

When initialized with a well trained separation model using feature recovery objective $(6)$, the permutation computation logic is not necessary for ASR objective, as shown in \eqref{eq:asr_loss}. For each training sample, we determine the label permutation by measuring the distance between $\mathbf{M}_{c} \odot |\mathbf{Y}|$ and spectrum of the reference signal $|\mathbf{X}_{c}|, c \in \{0, 1\}$. 
% As the decoder is auto-regressive, we can save half computation times of the decoder forward compared with the permutation version.
% Compared with the joint training from scratch, this method would
In this case we can save $50\%$ computation when calculating the loss values. Note that we found that directly optimizing with ASR objective from scratch leads to strong turbulence in model convergence and results in inferior performance, so we excluded this setup in the experiments.

\subsection{Model Compression}

Despite the promising results, the conformer based separation network usually employs model architecture with large parameter size. This usually brings difficulties in model deployment because of expensive computation and slow inference, especially on edge devices. 
To reduce the model size, we apply the recently advanced teacher student learning. Specifically, the well trained large separation network serves as the teacher and provides the reference separation for the smaller student model. 
The layer-wise TS objective $\mathcal{L}_{\text{LTS}}$ adds L2 distance between the hidden representation from teacher and student model in each layer, as shown in $(8)$:
\begin{equation}
      \mathcal{L}_{\text{LTS}} = \gamma_S \cdot \mathcal{L}_{\text{TS}} + \sum_{s} \gamma_{s} \cdot \left \Vert \mathbf{H}_s - \mathbf{H}_{g(s)}^\top \right \Vert_F,
\end{equation}
where $g(\cdot)$ is a uniform layer mapping function between indices from student layers to teacher layers and $\gamma_{s}$ is the weight value of the $s$-th hidden layer. $\mathcal{L}_{\text{TS}}$ removes the permutation computation to force the student network fully following the teacher's actions:
\begin{equation}
    \mathcal{L}_\text{TS} = \sum_c \left \Vert \mathcal{F}(\mathbf{M}_c \odot |\mathbf{Y}|) - \mathcal{F}(\mathbf{M}_c^\top \odot |\mathbf{Y}|) \right \Vert_F,
\end{equation}
where the $\mathbf{M}_{c}^\top$ refers to teacher's mask prediction of the speaker $c$.
% please add an ts equation for ts here
% $\gamma_{s} = (s + 1) \cdot (\sum_{i=0}^S (i + 1) + S + 1)^{-1}$

Moreover, we apply an \textit{objective shifting} (OS) \cite{chen2020recall} mechanism for more effective TS learning, which enables us to train the student network with both teacher's predictions and ground truth references. The objective function equipped with OS mechanism is:
\begin{equation}
    \mathcal{L}_{\text{OS}} = \omega_t \mathcal{L}_\text{FA} + (1 - \omega_t) \mathcal{L}_\text{LTS},
\end{equation}
where $\omega_t = \text{sigmoid}(-k(t-t_0))$ and $t$ refers to the training steps. $k$ and $t_0$ are hyper parameters. We describe the detailed exploration on separation with TS learning in a separate paper \cite{chen2021ultra}, and we refer audiences to that for more details.

% From our preliminary experiments, we also adopt the Objective Shifting (OS) mechanism to improve the performance of the TS learning, which enables us to train the student network with both teacher's predictions and ground truth references. The objective function equipped with OS mechanism is wrote as:
% \begin{equation}
%     \mathcal{L}_{\text{OS}} = \omega_t \mathcal{L}_\text{FA} + (1 - \omega_t) \mathcal{L}_\text{LTS}
% \end{equation}
% where $\omega_t = \text{sigmoid}(-k(t-t_0))$ and $t$ refers to the training steps. $k$ and $t_0$ are hyper parameters.

\section{Experimental Setup}

\subsection{Dataset}

We evaluate our methods on the LibriCSS \cite{chen2020continuous} dataset which consists of 10 hours of multi-speaker recording from a meeting room with an overlap ratio ranging from $0\%$ to $40\%$.  The recording device is a seven-channel circular microphone array and we use the signal of the first channel for the single channel performance evaluation. We evaluate our model in both utterance-wise and continuous evaluation as defined in \cite{chen2020continuous}.

For model training, we employ two datasets: the first one consists of 219-hour data using the close talk speech sampled from WSJ1 and the second one includes 1000-hour overlapped mixture whose source speech  is sampled from LibriSpeech. Both dataset are simulated according to the signal model described in equation \eqref{eq:signal_model} and each training sample contains one or two speakers. The artificial room impulse responses are generated using the image method.
For two-speaker cases, the mixing SDR ranges from -5 dB to 5 dB and four mixture types are considered following the work \cite{yoshioka2018multi}. Both directional and isotropic noises were added to each mixture with an SNR uniformly sampled between [0, 20] dB and [10, 20] dB, respectively. The directional noises are simulated by convolving the point source noise from MUSAN dataset with the RIRs.
% The artificial room impulse responses are  generated using the image method.

% \cite{panayotov2015librispeech} \cite{allen1979image}

\subsection{Separation Model}
% resulting in 26.03M parameters
% and finally results in 9.93M parameters

Following the description in Section 2, the Conformer structure is adopted as our primary network. The $\text{Cfmr}_\text{base}$ model has 16 encoder layers, 256 attention dimensions with 4 heads. The inner-layer of the FFN has 1024 dimensions and the kernel size and channel number used in CONV is 33 and 512, respectively. The $\text{Cfmr}_\text{small}$ model has the similar configurations but reduces the layer of the encoders to 6. 
% All the dropout layers in MHSA, FFN and residual connections are disabled to improve the performance.

We also evaluate a Transformer, a Convolution Recurrent Network (CRN) and a Dense-CRN network as representative baseline systems.
The Transformer model used here includes 16 encoder layers with the 4 head, 256 dimensional MHSA (using relative position encoding) and 1024 dimensional FFN. The CRN structure is a real-valued version of the DCCRN \cite{hu2020dccrn}, and consists of a 7-layer encoder/decoder with 3 layer bidirectional LSTMs. For Dense-CRN, we insert the DenseNet block between the layers in CRN's encoder and decoder, similar to the structure used in \cite{wang2020multi}. The Dense-CRN consists of a 8-layer encoder/decoder with 2 layer BLSTMs.
% It consists of a encoder, a decoder and a recurrent network where the encoder and decoder are stacked by several (de-)convolutional blocks, performing the down-sampling and up-sampling along the frequency axis, respectively.

 The 25 ms frame size with the frame shift of 10 ms is used for feature generation. A 512-point FFT size and hamming window are used in (i)STFT, forming the 257-dimentional masks and spectrum. The log spectrogram with utterance-wise mean variance normalization is extracted as the input feature for all the separation models. We only consider mixture with at most two speakers in experiments. The $\text{sigmoid}(\cdot)$ function is applied to final layer to make sure the masks have the value between 0 and 1.

\subsection{ASR Model}

We report our single channel results on two ASR back-end models. Both of them are trained on 960 hours of LibriSpeech training data, using the word piece units of the transcription as target. 
The first ASR, named $\text{ASR}_\text{matched}$,
is the ASR model used for ASR-oriented training of speech separation  following \eqref{eq:asr_loss}.
It consists of 12 Conformer \cite{chen2020conformer} encoder layers and 6 Transformer decoder layers with 80-dimentional log fbank as the input feature. This model shows WERs of $2.80\%$ and $6.80\%$ on Librispeech \textit{test-clean} and \textit{test-other}, respectively. 
The second ASR is the one developed in \cite{wang2019semantic}, which we
call $\text{ASR}$ \cite{wang2019semantic}. A concatenation of the filter bank and pitch features are used as the input to this model, and it 
achieves WERs of $2.08\%$ and $4.95\%$ on \textit{test-clean} and \textit{test-other}, respectively. 
For both ASR systems, an external language model trained on LibriSpeech's text corpus is applied with shallow fusion. The beam size and other decoding hyper-parameters are tuned on LibriSpeech \textit{dev-other} set.

% The first one follows the previous work in \cite{chen2020conformer}, using an open source E2E Transformer \cite{wang2019semantic} implementation which achieves WERs of $2.08\%$ and $4.95\%$ on \textit{test-clean} and \textit{test-other} set from LibriSpeech, respectively. The concatenation of the filter bank and pitch features are used as the input to this model. 
% The other one is used for tuning the separation model following \eqref{eq:asr_loss}. The model consists of  12 Conformer encoder layers and 6 Transformer decoder layers with 80-dimentional log fbank as the input feature. The performance of this model is slightly worse than \cite{wang2019semantic} and gives $2.80\%$ and $6.80\%$ WER on Librispeech test sets. During decoding, the shallow fusion with an external language model (LM) trained on LibriSpeech's text corpus is applied. The beam size and other decoding hyper-parameters are tuned on LibriSpeech "dev\_other" set.
% % We will use AM 1 and AM 2 to represent those two AMs in the experimental section.

\subsection{Training Details}

% In the experiments, we define training following $(6)$ as first stage training, and ASR-tuning, i.e. equation $(7)$ is referred as second stage training. As suggested in section 2.3, the network is randomly initialized for stage 1 training, and stage 2 is initialized with model trained from stage 1.

In our experiments, the training schemes are different depends on the model initialization. When the model is randomly initialized, the Transformer, $\text{Cfmr}_{\text{base}}$ and $\text{Cfmr}_{\text{small}}$ models are trained by $\mathcal{L}_\text{FA}$ with AdamW optimizer where the weight decay is set to $0.01$. A learning rate scheduler with linear warm-up and decay is used and the peak value of the learning rate is set to $10^{-4}$. The model is trained for 260k steps in total where the warm-up step is set as 10k. The CRN and Dense-CRN models are trained with the Adam optimizer with a weight decay of $10^{-5}$ and an initial learning rate of $10^{-3}$. Those networks are trained for a maximum of 260k steps and the learning rate is halved if no validation improvement is observed for two consecutive epochs. The early stopping strategy is applied to avoid over-fitting.

% When the model is random initialized, we use AdamW optimizer with the weight decay set to $0.01$. A learning rate scheduler with linear warm-up and decay is used and the peak value of the learning rate is set to $10^{-4}$. We train the model for 260k steps in total where the warm-up step is set as 10k. 
%  The CRN and Dense-CRN are trained with Adam optimizer with a weight decay of $10^{-5}$ and a initial learning rate of $10^{-3}$. 
% The network is trained for a maximum of 260k steps and the learning rate will halve if no validation improvement was observed for two consecutive epochs. The early stopping strategy was applied to avoid over-fitting.

When using the ASR-oriented objective function $\mathcal{L}_\text{ASR}$ or retraining the model by $\mathcal{L}_\text{FA}$ with additional data, the network is initialized with the pre-trained models, and we continue the training for 100k steps. During the first 25k steps, the learning rate is set to $4\times10^{-5}$ and starts linear decay after that. The AdamW optimizer with a weight decay of $0.01$ is used. For $\mathcal{L}_\text{ASR}$, $\lambda$ is set to $0.2$ and the unigram label smoothing is involved for calculation of the cross-entropy loss, which are kept same as the training configuration of $\text{ASR}_\text{matched}$. We apply gradient accumulation of 4 steps to increase the batch size due to the memory issues. The linear mel transform is used as the function $\mathcal{F}(\cdot)$ in our experiments.

For  TS learning, the  training configuration is same with the random initialization training, e.g., the learning rate scheduler, optimizer and the total training steps. We use $t = 1.5\times10^5$ and $k = 5 \times 10^{-4}$ in $\mathcal{L}_\text{OS}$ and  $g(s) = \{2,5,8,11,14,15\}$ for $s \in \{0, \cdots, 5\}$ in $\mathcal{L}_\text{LTS}$. All the models are trained using our self-developing tools based on PyTorch.

% For TS learning experiments, we use $t = 1.5\times10^5$ and $k = 5 \times 10^{-4}$ in $\mathcal{L}_\text{OS}$ and  $g(s) = \{2,5,8,11,14,15\}$ for $s \in \{0, \cdots, 5\}$ in $\mathcal{L}_\text{LTS}$. 

\section{Evaluation Results}

\subsection{Architecture Comparison}
We compare the utterance-wise performance on different network architectures in Table 1. Here, all front-end models are trained with 219 hours WSJ1 dataset,
and $\text{ASR}_\text{matched}$ is used for the back-end. 
% The comparison of the objective function $\mathcal{L}_\text{SA}$ and $\mathcal{L}_\text{FA}$ is conducted on $\text{Cfmr}_{\text{base}}$ structure, and we can see the clear WER reduction, especially on higher overlap ratio sets that ranges from 20\% to 40\%. 
On Conformer architecture, a clear WER improvement is observed when comparing the feature-level and the signal-level objectives, especially for high overlapped conditions.
For example, WER on $30\%/40\%$ subsets are reduced from $13.4\%/15.1\%$ to $12.1\%/13.9\%$.
% After the architecture search works, we present results of Transformer, CRN and Dense-CRN for experimental comparison. 
%The conformer based model shows a performance advantage. 
We also observe that
both  $\text{Cfmr}_\text{base}$ and  $\text{Cfmr}_\text{small}$ significantly outperform other popular network architectures. 
Compared with $\text{Cfmr}_\text{base}$, smaller model $\text{Cfmr}_\text{small}$ showed a clear WER increase, indicating that the necessity of more efficient training scheme such as model compression.

\begin{table}[!t]
\setlength{\tabcolsep}{1pt}
% \footnotesize
% \setlength{\tabcolsep}{4pt}
\centering
\caption{WER (\%) for utterance-wise evaluation with different separation model architectures. $\text{ASR}_\text{matched}$ is used.} % The numbers denote WER (\%) on LibriCSS test sets
 \vspace{-3mm}
\begin{tabular}{c|c|c|p{15pt}<{\centering}|p{15pt}<{\centering}|p{15pt}<{\centering}|p{15pt}<{\centering}|p{15pt}<{\centering}|p{15pt}<{\centering}}
\toprule
\hline
\multirow{2}{*}{\textbf{Separation}} & \multirow{2}{*}{\textbf{\#Param}} &  \multirow{2}{*}{\textbf{Loss}} & \multicolumn{6}{c}{\textbf{Overlap Ratio (\%)}} \\
\cline{4-9}
& & & \textbf{0S} & \textbf{0L} & \textbf{10} & \textbf{20} & \textbf{30} & \textbf{40} \\
\hline
No separation & - & - & 6.0 & 5.5 & 12.2 & 20.3 & 28.7 & 38.3 \\
\hline
\multirow{2}{*}{ $\text{Cfmr}_\text{base}$} & \multirow{2}{*}{26.03M} &  $\mathcal{L}_\text{SA}$ & \textbf{5.7} & \textbf{5.3} & 7.7 & 10.7 & 13.4 & 15.1 \\ 
&  &  $\mathcal{L}_\text{FA}$ & 5.8 & 5.4 & \textbf{7.4} & \textbf{10.0} & \textbf{12.1} & \textbf{13.9} \\ 
\hline
$\text{Cfmr}_\text{small}$ & 9.97M &  \multirow{4}{*}{$\mathcal{L}_\text{FA}$} &  6.0 & 5.3 & 8.1 & 11.0 & 13.4 & 15.3 \\
Transformer & 12.97M & & 5.7 & 5.4 & 8.3 & 12.0 & 15.0 & 17.4 \\
CRN & 20.37M &  & 6.1 & 5.8 & 8.4 & 12.5 & 16.5 & 19.3 \\
Dense-CRN & 17.99M & & 5.9 & 5.7 & 8.2 & 11.6 & 14.6 & 17.3 \\
\hline
\bottomrule
\end{tabular}
 \vspace{-3mm}
\end{table} 

% Besides, all the networks shown in the table trained with $\mathcal{L}_\text{FA}$ can surpass the first result on the Conformer structure released in \cite{chen2020conformer} while using the same AM (AM 2) for decoding.

% As the feature approximation function $\mathcal{L}_\text{FA}$ is designed to match the fbank feature that AM 1 uses, obviously we achieve more gains on it, e.g., WER on $30\%/40\%$ subsets reduce from $13.4/15.1$ to $12.1/13.9$ while on AM 2, $10.9/12.7$$\to$$10.1/12.1$.
% E2E \cite{wang2019semantic} $\text{E2E}_{\text{joint}}$

\begin{table}[!t]
\setlength{\tabcolsep}{0.5pt}
% \footnotesize
% \setlength{\tabcolsep}{4pt}
\centering
\caption{WER (\%) for utterance-wise evaluation with different training objective function and ASR back-end.}
 \vspace{-3mm}
\begin{tabular}{c|c|c|p{15pt}<{\centering}|p{15pt}<{\centering}|p{15pt}<{\centering}|p{15pt}<{\centering}|p{15pt}<{\centering}|p{15pt}<{\centering}}
\toprule
\hline
\multirow{2}{*}{\textbf{ASR}} & \multirow{2}{*}{\textbf{Separation}} &  \multirow{2}{*}{\textbf{Loss}} & \multicolumn{6}{c}{\textbf{Overlap Ratio (\%)}} \\
\cline{4-9}
& & & \textbf{0S} & \textbf{0L} & \textbf{10} & \textbf{20} & \textbf{30} & \textbf{40} \\
\hline
\multirow{3}{*}{$\text{ASR}_\text{matched}$} & \multirow{3}{*}{ $\text{Cfmr}_\text{base}$} & $\mathcal{L}_\text{FA}$ & 5.4 & 5.0 & 6.8 & 8.9 & 10.4 & 11.7 \\ 
&  & $\mathcal{L}_\text{ASR}$ & 5.1 & 4.7 & 6.4 & 8.5 & 9.8 & 10.7 \\ 
&  & $\mathcal{L}_\text{FA}$$\to$$\mathcal{L}_\text{ASR}$ & \textbf{4.8} & \textbf{4.6} & \textbf{6.1} & \textbf{8.0} & \textbf{9.2} & \textbf{10.0} \\ 
\hline
\multirow{4}{*}{$\text{ASR} \cite{wang2019semantic}$} & \multirow{3}{*}{ $\text{Cfmr}_\text{base}$}  & $\mathcal{L}_\text{FA}$ & 3.7 & 3.8 & 5.0 & 6.9 & 8.6 & 9.8 \\ 
&  &  $\mathcal{L}_\text{ASR}$ & 3.7 & 3.7 & 5.1 & 6.4 & 8.1 & 9.2 \\ 
&  & $\mathcal{L}_\text{FA}$$\to$$\mathcal{L}_\text{ASR}$& \textbf{3.6} & \textbf{3.6} & \textbf{4.7} & \textbf{6.4} & \textbf{7.7} & \textbf{8.4} \\ 
\cline{2-9}
& \cite{chen2020conformer} & $\mathcal{L}_\text{SA}$ & 5.4 & 5.0 & 7.5 & 10.7 & 13.8 & 17.1 \\
\hline
\cite{wang2020multi} & SISO$_{1+3}$ & \cite{wang2020multi} & 4.9 & 5.1 & 6.7 & 9.4 & 12.7 & 15.5 \\
\hline
\bottomrule
\end{tabular}
 \vspace{-3mm}
\end{table}  

\subsection{Comparison of Training Objective}

% We perform the stage 2 training by using the objective function $\mathcal{L}_{\text{ASR}}$ for the network optimization. As the training data of the stage 2 is different from stage 1, to remove the potential impact caused by the data difference, we also use $\mathcal{L}_\text{FA}$ for second round training. 

The comparison among models trained with different objective functions is shown in Table 2. 
Here, all models are trained on the 1000-hour LibriSpeech mixture data starting from the
pre-trained model with 219-hour WSJ1 mixture. 
The first and fourth rows indicate the results of the speech separation model retrained by $\mathcal{L}_\text{FA}$ while the second and fifth rows indicate the results of the model retrained by $\mathcal{L}_\text{ASR}$.
Finally, the rows with a loss $\mathcal{L}_\text{FA}$$\to$$\mathcal{L}_\text{ASR}$ refers the model using ASR-oriented objective, but initialized with the model corresponding to the first row. 

% In Table 2, $\mathcal{L}_\text{ASR}$ refers the stage 2 training described in section 3.4.
% $\mathcal{L}_\text{FA}$ refers the model trained with feature level objective $(6)$ using all 1219 hours data.  And $\mathcal{L}_\text{FA}$$\to$$\mathcal{L}_\text{ASR}$ refers the stage 2 model initialized with 1219 hours trained  
% This was done by concatenating the AM after the pre-trained separation model in stage 1 and passing the feature sequence of the separated speaker $\mathbf{X}'_c$ to the following ASR encoder and decoder.

Compared with Table 1, a clear benefit of using additional training data can be observed. The model retrained by $\mathcal{L}_\text{FA}$ reduces the average WER from $9.1\%$ to $8.0\%$ on $\text{ASR}_\text{matched}$ and the model trained by $\mathcal{L}_\text{ASR}$ demonstrates advantageous performance than $\mathcal{L}_\text{FA}$ on all subsets, showing the effectiveness of ASR-oriented optimization. 
Meanwhile, another $5\%$ WER reduction is observed by better initialization ($\mathcal{L}_\text{FA}$$\to$$\mathcal{L}_\text{ASR}$). 
We can also see that, even though the separation model is tuned with $\text{ASR}_\text{matched}$, the consistent performance improvement still exists when the ASR model is changed to ASR \cite{wang2019semantic} which uses a slightly different feature and architecture.
Finally, compared to the state-of-the-art system reported in \cite{wang2020multi}, our model shows a significant performance improvement, reducing the average WER from $9.1\%$ to $5.7\%$.

% Compared with the WER reported in Table 1, we observed that with such a two-stage training scheme, both two candidate loss functions, i.e., $\mathcal{L}_\text{FA}$ and $\mathcal{L}_\text{ASR}$ yield significant WER reduction on our ASR model. The gains on $\mathcal{L}_\text{FA}$ demonstrates a well initialization of the network and increasing the diversity of training data could contribute to a better result on single channel processing. Optimization with $\mathcal{L}_\text{ASR}$ gives additional improvement over all the six subsets on $\text{Cfmr}_\text{base}$, showing the effectiveness of the ASR fine tuning. The absolute WER reduction given by  $\mathcal{L}_\text{ASR}$ on $30\%/40\%$ overlapping subsets is $2.3\%/3.2\%$ compared with stage 1 and $0.6\%/1.0\%$ compared with $\mathcal{L}_\text{FA}$ of stage 2 results. We can also see that although the separation model is fine tuned with the our own ASR backend, the consistent performance improvement still exists when decoding using the ASR model \cite{wang2019semantic}, which uses a slightly different feature (fbank plus pitch) and architecture setups. We can further achieve better results by using a better initialization model for $\mathcal{L}_\text{ASR}$ training, i.e., $\mathcal{L}_\text{FA}$$\to$$\mathcal{L}_\text{ASR}$ in the Table 2, which illustrates the benefits of the ASR optimization.

\subsection{Model Compression}

The result for model comparison is shown in Table 3, where all models have $\text{Cfmr}_\text{small}$ architecture. $\mathcal{L}_\text{OS}$$\to$$\mathcal{L}_\text{ASR}$ means the model trained using ASR-oriented objective, with $\mathcal{L}_\text{OS}$ optimized model as the initialization. All the models are trained on 1000-hour Librispeech dataset.

Similar to Table 2, the two-stage training scheme significantly improves the WER of $\text{Cfmr}_\text{small}$.
Compared with a training from scratch, $\mathcal{L}_\text{OS}$ brings notable WER reduction and shows competitive results with the ASR-oriented training.
A further WER reduction is observed when the $\mathcal{L}_\text{OS}$ trained model is further fine tuned by the ASR-oriented objective function. 
After applying the TS learning, the performance gap between teacher and student models is reduced to $1.1\%$ WER increase, with $62\%$ reduction in parameter size. 

% Note that though $\text{Cfmr}_{\text{small}}$ has similar parameter size as CNN based system in \cite{wang2020multi}, it has essentially less computation, thus leading to faster inference.

% there is still the gap between $\text{Cfmr}_\text{small}$ and $\text{Cfmr}_\text{base}$, e.g., more than 1\% WER difference on higher overlapped subsets with ASR model \cite{wang2019semantic} compared Table 2 with 3. 
% The TS learning technique is deployed to achieve better results on the smaller models and the experimental results on model compression are shown in Table 3, where the stage 2 $\text{Cfmr}_\text{base}$ model trained with $\mathcal{L}_\text{ASR}$ is used as the teacher. %in the TS experiments.

% Compared with training from scratch, $\mathcal{L}_\text{OS}$ brings visible WER reduction on both two ASR models, even using the WSJ1 219 hours data for training, reducing the WER on $30\%/40\%$ subsets from $11.5\%/13.4\%$ to $10.4\%/11.7\%$. By using the TS learning result as the seed model, the second round ASR tuning brings better WER than the normal two stage training, e.g., $9.4\%/10.5\%$ $\to$ $8.7\%/9.5\%$ on higher overlapping sets. Meanwhile, this results can also beat the numbers of the previous work, which are shown in Table 2.

\begin{table}[!t]
\setlength{\tabcolsep}{2pt}
% \footnotesize
% \setlength{\tabcolsep}{4pt}
\centering
% ($\text{Cfmr}_\text{small}$)
\caption{WER (\%) for utterance-wise evaluation with $\text{Cfmr}_\text{small}$ trained by 
the TS learning and ASR-oriented objective function.}
 \vspace{-3mm}
\begin{tabular}{c|c|p{15pt}<{\centering}|p{15pt}<{\centering}|p{15pt}<{\centering}|p{15pt}<{\centering}|p{15pt}<{\centering}|p{15pt}<{\centering}}
\toprule
\hline
\multirow{2}{*}{\textbf{ASR}} &  \multirow{2}{*}{\textbf{Loss}} & \multicolumn{6}{c}{\textbf{Overlap Ratio (\%)}} \\
\cline{3-8}
& & \textbf{0S} & \textbf{0L} & \textbf{10} & \textbf{20} & \textbf{30} & \textbf{40} \\
\hline
% \multirow{3}{*}{\cite{wang2019semantic}} & % \multirow{2}{*}{$\mathcal{L}_\text{FA}$} & 4.1 & 4.1 & 6.1 & 8.6 & 11.5 & 13.4 \\ 
\multirow{4}{*}{$\text{ASR} \cite{wang2019semantic}$} & $\mathcal{L}_\text{FA}$ & 3.8 & 3.7 & 5.5 & 7.5 & 10.1 & 11.7 \\
 & $\mathcal{L}_\text{OS}$ & 3.8 & 3.7 & \textbf{5.3} & 7.1 & 9.5 & 10.7 \\
 & $\mathcal{L}_\text{ASR}$ & 3.9 & 3.9 & 5.7 & 7.7 & 9.4 & 10.5 \\
% & $\mathcal{L}_\text{OS}$ & 4.0 & 4.0 & 5.6 & 8.0 & 10.4 & 11.7 \\ 
 & $\mathcal{L}_\text{OS}$$\to$$\mathcal{L}_\text{ASR}$ & \textbf{3.7} & \textbf{3.8} & 5.4 & \textbf{7.0} & \textbf{8.7} & \textbf{9.5} \\ 
% \hline 
% % \multirow{3}{*}{Our} & $\mathcal{L}_\text{FA}$ & 6.0 & 5.3 & 8.1 & 11.0 & 13.4 & 15.3  \\ 
% % & $\mathcal{L}_\text{OS}$ & 5.6 & 5.2 & 7.5 & 10.1 & 12.1 & 13.7 \\
% % \cline{2-8}
% \multirow{2}{*}{Our} & $\mathcal{L}_\text{OS}$ & 5.4 & 5.0 & 7.0 & 9.7 & 11.3 & 12.7 \\
% & $\mathcal{L}_\text{OS}$$\to$$\mathcal{L}_\text{ASR}$ &  \textbf{5.1} & \textbf{4.8} & \textbf{6.5} & \textbf{8.5} & \textbf{10.2} & \textbf{11.1} \\
% Our & 1219h & $\mathcal{L}_\text{OS}$$\to$$\mathcal{L}_\text{ASR}$ & 5.1 & 4.8 & 6.5 & 8.5 & 10.2 & 11.1 \\
\hline
\bottomrule
\end{tabular}
 \vspace{-3mm}
\end{table}  

\subsection{Continuous Evaluation}

\begin{table}[!t]
\setlength{\tabcolsep}{1pt}
% \footnotesize
% \setlength{\tabcolsep}{4pt}
\centering
\caption{WER (\%) for continuous evaluation. $\text{ASR} \cite{wang2019semantic}$ is used.}
 \vspace{-3mm}
\begin{tabular}{c|c|c|c|c|c|c|c}
\toprule
\hline
\multirow{2}{*}{\textbf{Separation}} &  \multirow{2}{*}{\textbf{Loss}} & \multicolumn{6}{c}{\textbf{Overlap Ratio (\%)}} \\
\cline{3-8}
& & \textbf{0S} & \textbf{0L} & \textbf{10} & \textbf{20} & \textbf{30} & \textbf{40} \\
\hline
\cite{chen2020conformer} & $\mathcal{L}_\text{SA}$ & \textbf{6.9} & \textbf{6.1} & 9.1 & 12.5 & 16.7 & 19.3 \\
% Tfmr &  & $\mathcal{L}_\text{FA}$ & 11.6/13.3/13.3/14.4/18.0/19.2 \\
\hline
% \multirow{2}{*}{$\text{Cfmr}_\text{base}$} & 219h & $\mathcal{L}_\text{FA}$ & 10.8 & 11.8 & 11.5 & 13.6 & 16.9 & 18.0 \\
% \cline{2-9}
% & \multirow{3}{*}{1219h} & $\mathcal{L}_\text{FA}$ &  6.7/6.5/7.3/8.3/11.9/12.5 \\
\multirow{2}{*}{$\text{Cfmr}_\text{base}$} & $\mathcal{L}_\text{ASR}$ & 7.3 & 6.2 & 8.9 & 9.9 & 13.1 & 13.9 \\
&  $\mathcal{L}_\text{FA}$$\to$$\mathcal{L}_\text{ASR}$ & 7.5 & 6.4 & \textbf{8.4} & \textbf{9.4} & \textbf{12.4} & \textbf{13.2} \\
\hline
% \multirow{4}{*}{$\text{Cfmr}_\text{small}$} & \multirow{2}{*}{219h} & $\mathcal{L}_\text{FA}$ & 12.6 & 13.9 & 13.9 & 15.6 & 18.9 & 20.6 \\
% &  & $\mathcal{L}_\text{OS}$ & 12.9 & 13.9 & 13.6 & 15.5 & 18.0 & 18.8 \\
% \cline{2-9}
% & \multirow{3}{*}{1219h} & $\mathcal{L}_\text{FA}$ &  \\
% &  & $\mathcal{L}_\text{ASR}$ & \\
\multirow{2}{*}{$\text{Cfmr}_\text{small}$}  & $\mathcal{L}_\text{ASR}$ &  11.6 & 10.3 & 13.0 & 13.7 & 17.3 & 17.8 \\
&  $\mathcal{L}_\text{OS}$$\to$$\mathcal{L}_\text{ASR}$ & 8.2 & 6.5 & 10.0 & 11.2 & 14.5 & 15.6 \\
\hline
% \cite{wang2020multi} & 129h & 
% \hline
\bottomrule
\end{tabular}
 \vspace{-3mm}
\end{table} 

We follow the similar chunk-wise processing described in \cite{chen2020conformer} for continuous speech separation with signal-wise stitching. Compared with the previous work in \cite{chen2020conformer}, both $\text{Cfmr}_\text{base}$ and $\text{Cfmr}_\text{small}$ show remarkable improvement for the highly overlapping test sets after the training based on  $\mathcal{L}_\text{ASR}$, while the $\text{Cfmr}_\text{small}$ has a 5 times smaller parameter size compared with the 58.72M-parameter model of \cite{chen2020conformer}. However, the numbers on the non-overlapping sets has slight performance degradation compared with the results of \cite{chen2020continuous}, which suggests us the potential room of improvement under current framework.

% Within the same network architecture, the $\mathcal{L}_\text{ASR}$ training brings overwhelming WER results on all the six subsets compared with the $\mathcal{L}_\text{FA}$, which is consistent with the utterance-wise evaluation. 
% For  $\text{Cfmr}_\text{small}$, the WER after the TS training with OS mechanism ($\mathcal{L}_\text{OS}$) is also better than that of the $\mathcal{L}_\text{FA}$ while only using 219 hours training data. 

% A 2.4s sliding audio chunk (with a hop size of 0.8s) is used as input and a 0.8-s-long separated results are produced for each step. The first 1.2s and last 0.4s are treated as the past and future contexts, respectively. 
% with slight degradation for the non-overlapping sets

% With the default hybrid AM released in \cite{chen2020continuous}, the performance is still much better than the best number in \cite{chen2020conformer} while showing  worse results for the non-overlapped data compared with \cite{wang2020multi}. This may caused by the difference of the optimization objective between the hybrid and E2E AM, which uses senones and word pieces as the modeling unit, respectively. We will explore the method to further improve the continuous results in the future work.

% We use the AM 2 for decoding and the results of the continuous separation on several selected models are shown in Table 4.
% The overlapped region between the chunks are utilized to determine the permutation order between the neighboring time steps. 

\section{Conclusions}

In this paper, we investigate several practical aspects of single channel speech separation for ASR, including a two-stage training scheme to utilize both feature and ASR-oriented optimization criterion and the TS training as the final step to compress the model size. The conclusions on the utterance-wise and continuous evaluations are consistent and the performance gains from the ASR-oriented training could be shifted to another different ASR model. The state-of-the-art results using a smaller Conformer model with less than 10M parameters on LibriCSS dataset are achieved on utterance-wise evaluation, which gives an average $2.7\%$ absolute WER reduction compared with the best results shown before. For the continuous evaluation, we achieve an average relative WER improvement of $6.4\%$ with significant gains on overlapped sets. %while the non-overlapping sets has slight performance degradation.
% The competitive WER on continuous evaluation is reported, while .

\newpage

\bibliographystyle{IEEEtran}

\bibliography{mybib}

\end{document}